\documentclass{ws-rv975x65}
\usepackage{ws-rv-thm}     
\usepackage{ws-rv-van}     
\usepackage{phonetic}
\usepackage{subfigure}
\usepackage{wrapfig}
\let\captionbox=\undefined
\usepackage{caption}
\makeindex

\begin{document}

\chapter[Radio Detection of High Energy Neutrinos]{Radio Detection of High Energy Neutrinos}

\author[]{Amy L. Connolly}
\address{Ohio State University, Department of Physics and CCAPP, Columbus, OH 43210 USA}
\author[A. Connolly and A. G. Vieregg]{Abigail G. Vieregg}
\address{University of Chicago, Department of Physics, Enrico Fermi Institute, Kavli Institute for Cosmological Physics, Chicago, IL 60637 USA}

\body

\section{Introduction}
Ultra-high energy (UHE) neutrino astrophysics sits at 
the crossroads of particle physics, astronomy, and astrophysics.\footnote{Here, we loosely define the UHE regime to
be the energy range near $10^{18}-10^{21}$~eV.  A few radio experiments have sensitivities that reach to lower energies,
while others have energy thresholds at higher energies.}
Through neutrino astrophysics, we can uniquely explore the structure and evolution of the 
universe at the highest energies at cosmic distances 
and test our understanding of particle physics at energies
greater than those available at particle colliders.  

The detection of UHE neutrinos would shed light on the nature of the 
astrophysical sources that produce the highest energy particles in the universe.
 Astrophysical sources almost certainly produce UHE neutrinos in hadronic processes.  
Also, neutrinos above $10^{17}$ eV should be produced through the GZK effect~\cite{g,zk}, where
extragalactic 
cosmic rays above $10^{19.5}$~eV interact with the cosmic microwave background within tens of
Mpc of their source.   It was first pointed out by Berezinsky and Zatsepin~\cite{Berezinsky:1969zz, Berezinsky:1970}
that this cosmogenic neutrino flux, sometimes called ``BZ'' neutrinos, could be observable.
These neutrinos would point close to the cosmic ray production site both because of the close proximity of the
interaction to the source and because the decay products are Lorentz boosted along the line of sight.
The latter effect constrains the direction the most strongly ($1~\rm{MeV}/1~\rm{EeV}=10^{-12}$ whereas $100~\rm{Mpc}/1~\rm{Gpc}=0.1$).
Since cosmic rays do not follow straight paths in magnetic fields and get attenuated above the GZK threshold, and
high energy photons (E$>10^{14}$~eV) are attenuated by the cosmic infrared background,
neutrinos offer the unique ability to point to the location of the
highest energy cosmic accelerators in the sky.

UHE neutrino measurements will
also have important 
implications for high energy particle 
physics and determining neutrino properties.  A sample of UHE neutrinos would allow for
a measurement of the $\nu$-p cross
section~\cite{Connolly:2011vc,Klein:2013xoa} and
 a direct test of weak interaction couplings 
at center-of-mass (CM)
energies beyond the Large Hadron Collider 
(a 10$^{18}$~eV neutrino collides with a proton at rest at 45~TeV CM).  A strong
constraint on the UHE neutrino flux can even shed light on models of Lorentz invariance violation~\cite{anitaLorentz,Anchordoqui:2014hua}.  

IceCube has recently discovered neutrinos with energies up to a few $10^{15}$~eV
 that are likely produced by astrophysical
sources directly~\cite{bertErnie,bigBird}.  Therefore,
there is now a pressing motivation to measure  
the energy spectrum above $10^{15}$~eV with improved sensitivity, to
provide insight into the 
origin of the seemingly cosmic events and the particle acceleration mechanisms that give rise to them and to 
determine the high-energy extent of the spectrum. 
 
\begin{wrapfigure}{r}{0.65\textwidth }
          \begin{center}\includegraphics[width=0.60\textwidth]{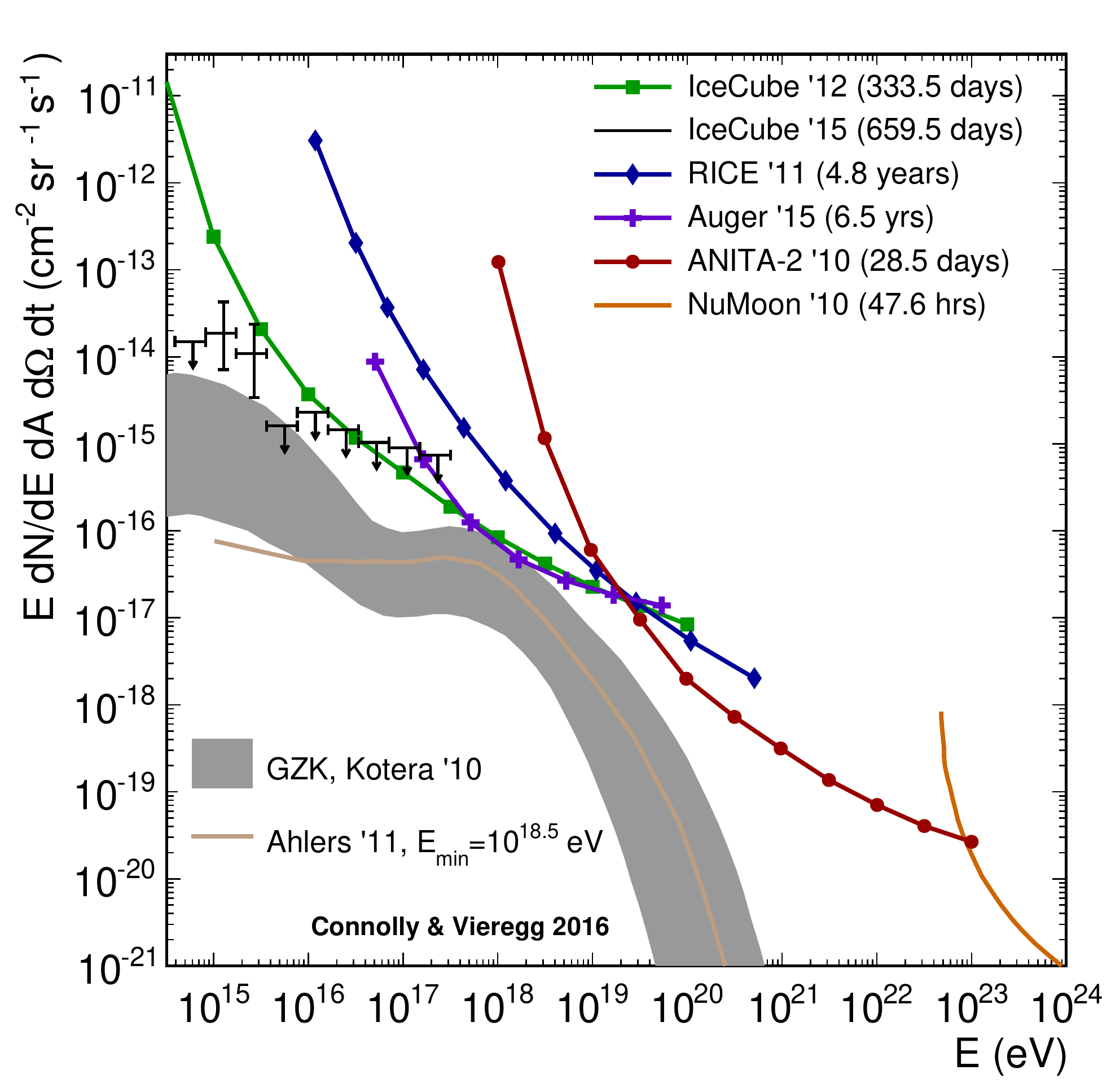}
           \end{center}
           \captionsetup{width=0.60\textwidth}
          \caption{The current most competitive experimental constraints on the all-flavor 
          diffuse flux of the highest energy neutrinos compared to representative model predictions~\cite{ahlers,kotera}.  
          Limits are from IceCube, Auger, 
            RICE, ANITA and NuMoon
            ~\cite{augerLimit,rice,anita2,anita2err,icecube2015,buitinkconstraints2010}.  
            Also shown
            is the astrophysical neutrino flux measured by IceCube~\cite{icecube2015}.
          \label{fig:limit}}
        \end{wrapfigure}

Figure~\ref{fig:limit} shows the current best limits on the high energy diffuse neutrino flux compared with
 a variety of GZK production models.  
Beyond the astrophysical neutrino events measured up to $\sim10^{15}$~eV, 
IceCube sets the best  limits on the high energy neutrino flux up to $10^{17.5}$~eV, 
and now shares the claim for the best constraints in the region
up to $10^{19.5}$~eV with the Pierre Auger Observatory (Auger)~\cite{icecube2015,augerLimit}.
The current best limit on the flux of neutrinos
above $10^{19.5}$~eV comes from the Antarctic Impulsive Transient Antenna (ANITA) 
experiment~\cite{anita2,anita2err}. 

Note that on the vertical axis of Figure~\ref{fig:limit} is the differential flux $dN/dE$ multiplied by one
power of $E$; the product is proportional to $dN/d\log_{10}E$.  On this plot, an experiment that increases
its sensitivity in an energy-independent way, for example by increasing its live time, will have its limits 
move only downward.  An experiment that decreases its energy threshold with no other change 
will have its constraints move
only to the left on this plot.

Despite the competitive limits currently imposed by
optical detection experiments, it would be prohibitively costly to build a detector utilizing the optical signature
that would be sensitive to
the full range of predicted possible cosmogenic neutrino populations above $10^{17}$~eV, due
to the detector spacings set by the absorption and scattering lengths of optical light in ice~\cite{Ackermann:2006pva}.
IceCube-Gen2 is a proposed IceCube expansion to 100-300~km$^2$ scale for the detection of
neutrinos above $10^{13.5}$~eV, focusing on energies above which the atmospheric neutrino background is not
overwhelmingly dominant, and with
 the discovery potential for BZ neutrinos~\cite{Aartsen:2014njl}.
However, to achieve sensitivity to the full range of BZ neutrino models,
we must instead turn to a different detection mechanism 
that allows us to instrument larger volumes for comparable cost.
Neutrino telescopes that utilize the radio 
detection technique search for the coherent, impulsive radio signals that are emitted 
by electromagnetic particle cascades induced by neutrinos interacting with a dielectric.  
Radio UHE neutrino detection requires a volume $\mathcal{O}(100)$ km$^3$ of dielectric material, which 
limits the detection medium to be
naturally-occurring, that allows radio signals to pass through without significant attenuation over
lengths $\mathcal{O}(1)$~km.  Current and proposed 
experimental efforts in this field monitor or plan to monitor immense volumes of glacial ice, whose radio 
attenuation properties have been directly measured at multiple locations in Antarctica and Greenland, and have 
the desired clarity~\cite{Barrella:2010vs,besson,barwickSouthPole,araWhitepaper,avva,Barwick_Berg_Besson_Duffin_2014}.
We note that although IceCube-Gen2 is nominally on an expansion of the array of optical sensors, a 
radio array component may be considered for enhancement of sensitivity in the energy range~$10^{16}$-$10^{20}$~eV~\cite{Aartsen:2014njl}.

Even some of the most pessimistic models
predicting BZ neutrino fluxes are within reach of planned experiments using the radio technique. 
These more pessimistic models tend to have heavy cosmic-ray composition or
a weak dependence of source densities on redshift, within
constraints set by other measurements.    
Recent measurements with the Auger and the Telescope Array 
disfavor a significant iron fraction in the cosmic-ray composition at energies up to 
and even exceeding $10^{19.5}$~eV, which in
turn favors higher fluxes of BZ neutrinos than if the highest energy cosmic rays were pure 
iron~\cite{auger_2014_composition,telescopeArray}.  
An experiment that has a factor of $\sim$50 improvement 
over the best sensitivity currently achieved by IceCube near $10^{18}-10^{19}$~eV, or
reduces the energy threshold with an ANITA-level sensitivity by about a factor of $\sim$50
would reach these pessimistic neutrino flux expectations.
For most models, such an experiment would
observe enough events to make an important impact in our understanding of UHE
astrophysics and particle physics utilizing the highest energy observable particles in the universe.
We note, however, that even these pessimistic models can be
evaded through alternate explanations for the cosmic ray data, or more exotic scenarios~\cite{Unger:2015laa,Globus:2015xga,anitaLorentz}.

\section{Motivation for the Radio Technique}
	\subsection{Askaryan Effect} 
	In 1962, physicist Gurgen Askaryan predicted 
	that an energetic electromagnetic cascade in a dense dielectric medium 
	should produce observable, coherent electromagnetic radiation~\cite{askaryan}.  
	When an energetic charged particle or photon produces an electromagnetic cascade in a dielectric medium, 
	the shower will
	acquire a $\sim20$\% negative charge excess.   
        This comes about primarily through Compton scattering of electrons in the medium, 
	but also from the annihilation of positrons in the shower with electrons in the medium.  If the charge excess 
        is moving at 
	a velocity greater than the phase velocity of light in the medium, the medium will emit Cherenkov radiation.   
        The radiation is emitted most strongly at the angle given by
        $\cos{\theta}=1/n$ with respect to the shower
	axis, and the power radiated at optical wavelengths is proportional to the shower energy.  In ice, a common
        medium for detection of neutrino interactions, $n=1.78$ at radio frequencies, giving $\theta=57^{\circ}$.    
        For wavelengths longer than the 
        size of the shower along the transverse dimension (perpendicular
        to the shower axis), the signal is emitted coherently and the electric field strength
        is proportional to the shower energy.  The Cherenkov radiation is coherent at wavelengths longer
        than the Moli\`{e}re radius in a dense medium ($\sim10$~cm), 
        corresponding to frequencies~$\lesssim$1~GHz.

\begin{figure}
            \centering{
            	\includegraphics[width=0.47\textwidth]{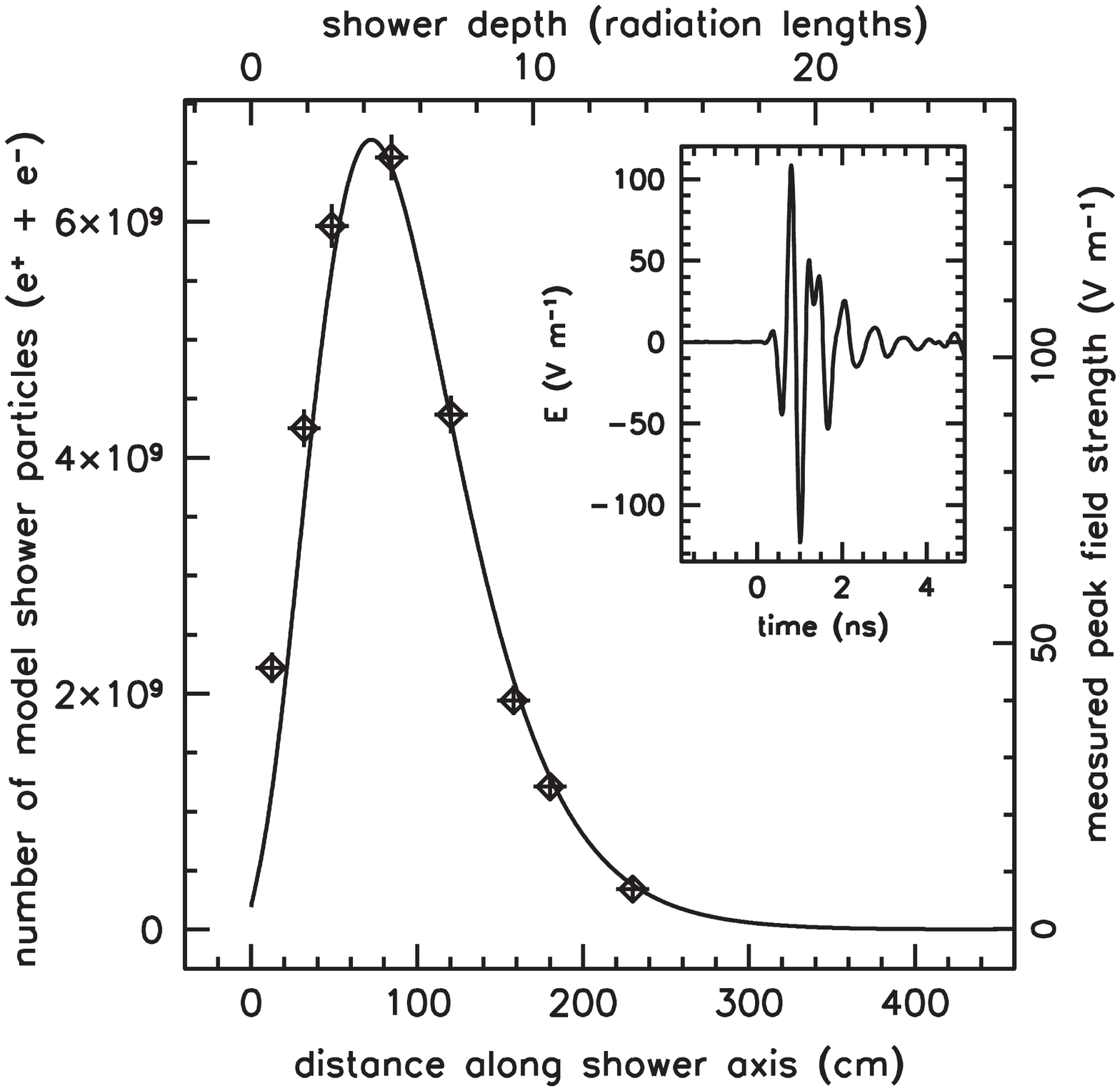}\hspace{0.1in}
    		\includegraphics[width=0.46\textwidth]{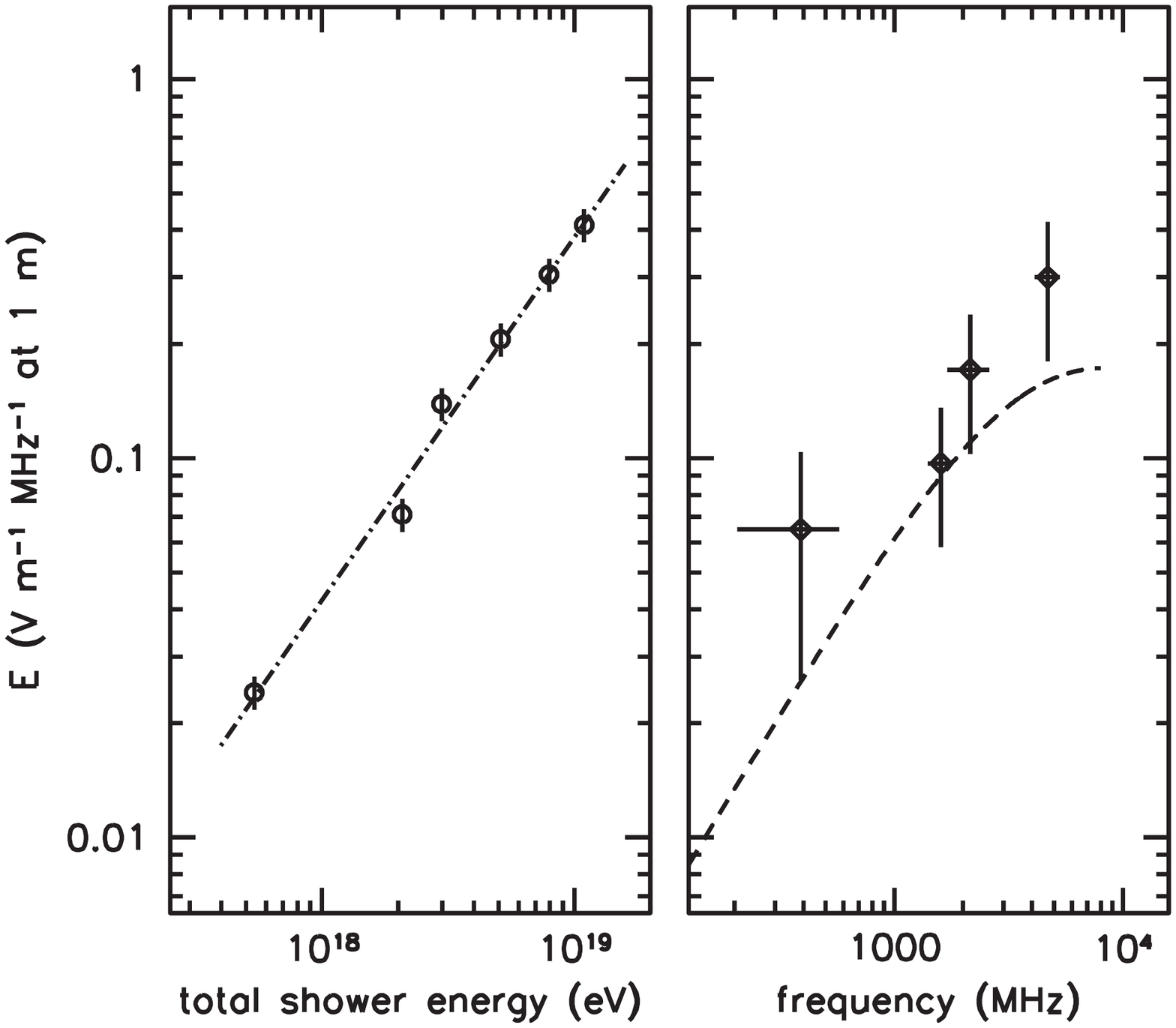}}
          
          \caption{The first observation of the Askaryan effect, from Reference~\cite{saltzbergSand}.  (a) The 
          field strengths closely follow the expected shower profile along the shower axis.  The time-dependent
           electric field  measured near shower maximum is shown in the inset.  
           (b) The measured electric field plotted against
            shower energy.  The dashed line shows a least-squares fit to the data, consistent with complete
            coherence.
            (c) The measured electric field spectrum compared to a semi-empirical parametrization is shown as a dashed line.}
          \label{fig:askaryanSand}
        \end{figure}

        The Askaryan effect was first observed in a beam test at SLAC National Accelerator Laboratory 
        in 2001~\cite{saltzbergSand}.  Using a target
        of silica sand to produce showers induced by brehmsstrahlung photons from a 28.5 GeV electron beam 
        and a series of broadband,
        microwave horn antennas with bandwidths covering the range 0.3 to 6 GHz,
         the beam test showed the first experimental
        evidence of the coherence effect predicted by Askaryan in the 1960's.  
        The results of the beam test are shown in Figure~\ref{fig:askaryanSand}~\cite{saltzbergSand}.
        Figure~\ref{fig:askaryanSand}a shows the measured shower profile as a function of
        distance along the shower axis, with an inset of the time-domain electric field measured near shower maximum.  
        The measured shower profile (diamonds) matches the prediction well, and the signal is broadband and impulsive.
        Figure~\ref{fig:askaryanSand}b shows the measured electric field strength as a function
        of shower energy, controlled by changing the number of photons that initiated the shower. 
  	The linear dependence of the electric field on the energy of the shower is evidence for coherence.
        Figure~\ref{fig:askaryanSand}c shows the measured electric
        field strength as a function of frequency compared to results from a Monte Carlo simulation.

%

        The observation of the Askaryan effect was confirmed 
        later in rock salt and again in ice at separate subsequent beam tests at  
        SLAC in 2004 and 
        2006, respectively~\cite{gorhamSalt, anitaIce}.  
        The observation of the Askaryan effect in ice is of particular note, since many
        experimental efforts monitor large volumes of naturally occurring ice.
        Although the lunar regolith (sand) continues to be used as a target for neutrino searches~\cite{westerbork,Bray:2013ta} and rock salt has been investigated as a potential target medium~\cite{salsa}, 
        most efforts use ice as a target due to its remarkable naturally-occurring volume, radio clarity, and uniformity.

        \subsection{A Radio Clear Dielectric for a Detector}
             To determine the suitability of different naturally-occurring
        dielectric media for the detection of radio emission from highly energetic neutrinos,
        many measurements have been made of the dielectric
        attenuation at radio frequencies at different sites around the world.   The field attenuation length $L_{\alpha}$ 
        is defined
        to be the distance over which the distance-corrected electric field ($E(r)$) 
        drops by a factor of $1-1/e$.  At a distance $r$:
        \begin{equation}
        rE(r)=E(0)e^{-r/L_{\alpha}} .
        \end{equation} 
        	The following equation given in~\cite{Barwick_Berg_Besson_Duffin_2014} 
	is useful for relating the field attenuation length reported in particle
	astrophysics to the dielectric loss per distance $N_L$ reported by geophysicists:
	\begin{equation}
	N_L \rm{[dB/km]} = 8686.0 \left<  L_{\alpha} [\rm{m}] \right>.
	\end{equation}

	Askaryan himself proposed dense media such as ice, salt, and even the lunar regolith for the detection of 
         cosmic ray showers~\cite{askaryan}.  Ice was later proposed as a 
        promising, cost-effective medium for the
	detection of showers produced by high energy neutrino interactions as well~\cite{zheleznykh}.  	
	Several measurements have confirmed that there are naturally-occurring ice sheets
	that exhibit
        the dielectric properties necessary to make
	an excellent medium for neutrino detection.
        Measurements in naturally-occurring rock salt have found more modest attenuation lengths, and 
        the lunar regolith has also been used as a target of observation for  neutrino interactions above $\sim10^{21}$~eV.
   
        \subsubsection{Attenuation Length of Glacial Ice}
        \label{sec:attenuation_ice}
        One of the most promising sites for a large radio detector 
        is in the ice near the South Pole, which sits atop $\sim2.8$~km of radio-clear glacial ice.
        The measurement of the radio attenuation length in ice at the South Pole with the smallest uncertainties
        and largest horizontal component~\cite{barwickSouthPole, araWhitepaper}
        gives a depth-averaged field attenuation length $\langle L_\alpha \rangle$ at 300~MHz 
        of $1600^{+255}_{-120}$~m over the top 1500~m of
        ice~\cite{araWhitepaper}, the depth over which a surface or sub-surface experiment 
        would be the most sensitive to neutrino interactions. This result is derived from a comparison between 
        the amplitude of a broadband signal transmitted from an antenna 
        deployed on an IceCube string 2500~m below the surface and 
        the amplitude of the same signal received at an
        antenna buried at about 30~m below the surface, 2000~m away
        horizontally.  
              
        A site with similar properties is found at the peak of the 
        the high plateau in Greenland, at Summit Station, which sits on top of 3~km of glacial ice.
        The attenuation length of the ice at Summit Station was measured by 
        bouncing an impulsive radio signal off of the bottom of the glacier,
        measured at $3014^{+48}_{-50}$~m depth, and comparing the strength of the return signal to a direct measurement
        between the transmitter and the receiver through air.
        The field attenuation length has been measured to be 
        $\langle L_\alpha \rangle=947^{+92}_{-85}$~m at 75~MHz averaged over all depths~\cite{avva}.  
        The power reflection coefficient $R$ is assumed to be 0.3 for 
        the ice-bedrock interface.
        For a reasonable extrapolation to higher frequencies, this 
        measurement is consistent with measurements in Reference~\cite{macgregor} reported at 150-195~MHz.
        Accounting for a depth-dependent temperature profile and extrapolating
        using a frequency dependence of -0.55~m/MHz, based on an ensemble of previous measurements,
        yields a field attenuation length of      
        $\langle L_\alpha \rangle=1022^{+230}_{-253}$~m over the top 1500~m of ice at 300~MHz~\cite{avva}.

        Another promising site is Moore's Bay on the Ross Ice Shelf, 
        where the average measured depth of the ice is $576 \pm 8$~m~\cite{Barwick_Berg_Besson_Duffin_2014}.
        Using a technique where reflection loss (found to be -1.7~dB) 
        is separated from attenuation loss by separating
        the transmitter and receiver by over 500~m at the surface, the measured 
        depth-averaged field attenuation length $\langle L_\alpha \rangle$ between 100 and 850~MHz is 
        $\left( 460\pm20\right) - \left(180 \pm 40 \right) \nu $~m, where $\nu$ is the frequency of interest
        in~GHz.  This corresponds to a 
        frequency dependence of -0.18~m/MHz~\cite{Barwick_Berg_Besson_Duffin_2014,Barrella:2010vs}. 
        The shorter attenuation length at Moore's Bay compared to 
        potential deep sites is due primarily to the fact that the ice is warmer at Moore's Bay.

        \subsubsection{Attenuation Length of Rock Salt}
	Large, naturally-occurring deposits of rock salt have also been investigated as a possible detection
        medium for radio emission from  neutrino interactions.  Investigations at three different locations in the United States 
        (WIPP in New Mexico,
        Hockley in Texas, and the Cote Blanche mine operated by the North American Salt Company in St. Mary Parish, Louisiana)
        ~\cite{hockley,coteBlanche} indicate that the attenuation length of rock salt is heavily dependent on the 
        moisture content, layering, and composition of the salt.  The longest attenuation length was measured directly at Cote 
        Blanche by transmitting impulsive, broadband radio signals directly through the salt from a 
        transmitter to a receiver with bandwidths spanning 
        125 to 900 MHz in boreholes drilled up to 200~ft. into the salt.
        
        The field attenuation length  $\langle L_\alpha \rangle$ at 300 MHz was measured to be $63\pm3$~m at Cote Blanche, 
        which is significantly shorter
        than measurements of glacial ice have shown.  However, the relatively high density of rock salt compared to ice boosts
        the neutrino interaction cross section by a factor of about 2.5, which would make up for some of the 
        volumetric loss due to the 
        shorter radio attenuation length.  In addition, 
        in salt the Cherenkov emission would be broader in solid angle~\cite{jaime_media}, leading to an increase in
        achievable effective volume,
        and an overburden of bedrock would block backgrounds from above such as galactic noise and cosmic rays.

        \subsubsection{Attenuation Length of Lunar Regolith}
        The lunar regolith is another possible radio-transparent target for interactions of highly energetic neutrinos. 
        The radio attenuation
        length of samples that have been returned to Earth has been shown to be $\mathcal{O}(20)$~m at 1~GHz 
        and $\mathcal{O}(200)$~m at 
        100~MHz~\cite{olohoeft}, comparable to that of rock salt. 
        However, these values were found to vary across samples, depending on their exact composition. 
        Although estimates of the depth of the moon's regolith vary widely, and is expected to vary over the lunar
        surface, conservative depth estimates
        are near 10~m~\cite{james_unanswered}.  Sensitivity estimates for the Goldstone Lunar Ultra-high-energy Neutrino Experiment
        (GLUE) experiment assume a 10~m regolith depth,
        while estimates for the NuMoon experiment assume a depth of up to 500~m~\cite{glue,Scholten_Bacelar_Braun_Bruyn_2006}.
        
\section{Experimental Strategies}

\subsection{Assessing the Sensitivity of an Experiment}
\label{sec:sensitivity}
There are a few factors that determine the number of neutrinos expected to be detected from a given experiment, 
and these are the factors that go into an experimental design.
We often define an energy-dependent, water-equivalent effective volume$\times$solid angle $\left[V\Omega\right]_{\rm{eff}} (E)$ 
for an experiment, where $E$ is the neutrino energy, and it is given by 
\begin{equation}
\left[ V\Omega\right]_{\rm{eff}} (E) = V \cdot 4\pi \cdot \varepsilon_V (E) \cdot \rho_{\rm{H20}}/\rho_{\rm{det}}, 
\end{equation}
where $V$ is the total volume of the detection medium, $\varepsilon_V (E)$ is the fraction of neutrinos interacting in that volume that pass the trigger, and $\rho_{\rm{det}}/\rho_{\rm{H20}}$ is the density of the detection medium relative to water.  In order to predict the number of neutrino events $N$ detected in an experiment 
from a given flux model $F(E)=dN/dE/d\Omega/dt$, we need to 
define an effective area$\times$solid angle $ \left[A\Omega\right]_{\rm{eff}} (E) $ so that 
\begin{equation}
N=\int F(E) \cdot \left[ A\Omega\right]_{\rm{eff}} (E)  \cdot T \cdot dE, 
\label{eq:N}
\end{equation}
where $T$ is the livetime of the experiment.  If the thin-target 
approximation is valid, i.e., the dimensions of the detection medium are much smaller than the interaction length $\ell (E)$,
then we can take $\left[ A\Omega_{\rm{eff}}\right] (E) = \left[ V\Omega\right]_{\rm{eff}} (E) /  \ell (E)$.

The effective area typically increases with energy, and there is some energy below which the
experiment does not expect to see a significant number of events.  This is called the energy threshold, and 
is not well-defined quantitatively, 
but is set by the energy-dependent efficiencies at the trigger and analysis stages (analysis efficiencies are not
included in the above equations).  In turn, the 
trigger efficiencies are set by the trigger thresholds necessary to maintain  the rate of data acquisition that is possible
given the level of thermal noise that is seen.
Since the high-energy
neutrino flux (from GZK models and direct production models) is a falling spectrum at high energies, 
the event rate can be strongly dependent on the energy threshold.

\subsection{Experimental Approaches}
Although most experiments that aim to detect radio emission from high energy neutrino interactions are similar in their approach, 
determining the best experimental configuration is a tradeoff among a variety of factors.  All experiments that 
rely on observations of 
Askaryan emission need to monitor a large 
volume of radio-transparent dielectric.  However, many different approaches have been explored, such as
the observation of the Antarctic continent 
or Greenland from a balloon or satellite, ground-based techniques that deploy instruments directly on the 
surface or at depth surrounded by
large volumes of ice or salt, and observation of the lunar regolith from afar.  Other experiments seek the radio emission
produced by an air shower resulting from the decay products of a tau lepton produced from a neutrino interaction.
All have different benefits and drawbacks.

\subsubsection{Experiments Using the Askaryan Technique}
We can see the trade-off between balloon-borne experiments and detectors on the ground in terms of the
variables defined in Section~\ref{sec:sensitivity}.  Putting a detector farther from the neutrino interactions increases the energy threshold.
However, from high altitudes an experiment can view a larger area of ice compared to an experiment on
the ground, which increases  
$\left[A\Omega\right]_{\rm{eff}}$. 
In general, detectors
that observe large volumes from far distances (from a balloon, satellite, or viewing the moon) 
are the best probes of the flux of the highest energy neutrinos.  We will call these ``view-from-a-distance''
experiments. 
Balloon experiments such as ANITA~\cite{Gorham:2008dv}
and the proposed ExaVolt Antenna (EVA)~\cite{Gorham:2011mt}
fly at $\sim 37$~km altitude above the Antarctic ice sheet.
ANITA has set the strongest limits above $10^{19.5}$~eV.     Telescopes such as the Parkes Lunar Radio Cherenkov Experiment~\cite{James:2007rq}
 aimed at the moon are seeking Askaryan
emission from the sandy regolith at the lunar surface.
The PRIDE experiment is even exploring other worlds as possible detection media, such as
Enceladus or Europa, icy moons of Saturn and Jupiter, respectively~\cite{Pride}.

Detectors on the ground or embedded in their detection medium, such as the
Askaryan Radio Array (ARA) and The Antarctic Ross Ice Shelf Antenna Neutrino Array (ARIANNA), which sit in and on glacial ice,
have a lower energy threshold than ones that are situated farther from the interactions.
In addition,  the livetime can be much longer  for an embedded experiment
compared to a balloon experiment (years compared to several weeks).  
The drawback of being so close to the interactions is that a smaller volume of ice can be monitored by each detector
element compared to  from above,
which reduces $\left[ A\Omega\right]_{\rm{eff}}$ for the same number of detectors.  Therefore, embedded detectors consist of
an array of antennas covering a large area.

View-from-a-distance experiments and in-ice experiments utilizing the Askaryan technique
thus have different strengths for probing high energy astrophysics.
For example, experiments that view from far distances have the greatest potential to measure the high energy cutoff
of the astrophysical sources.
Experiments on the ground, due to their lower threshold, are necessary to reach the heart of the 
GZK-induced neutrino spectrum, predicted to be at about $10^{18}$~eV, and if thresholds can be reduced, 
could measure the astrophysical neutrino spectrum above the $\sim1$~PeV energies corresponding to the highest
energy events so far observed by IceCube. 

Once neutrino events are observed, embedded detectors also have the potential to achieve better angular resolution
for neutrinos.  By placing detectors
relatively far apart (tens of meters), one can in principle view the Cherenkov cone from different
angles, and use measured polarizations to reconstruct the direction of the shower and thus the neutrino direction.
This is unlike a balloon-borne experiment whose size is tightly constrained.
  The energy resolution is also improved with the ability to pinpoint the neutrino interaction location.  
  Although these techniques are not yet mature, their development will become of extreme importance as soon as the 
  first  neutrinos are measured using the radio technique.

A few experiments pioneered the radio technique to search for high energy neutrinos by searching for the
Askaryan signature in various media. 
Data from the FORTE~\cite{forte} satellite
was used to search for neutrino interactions in Greenland ice in the late 1990's.   The Parkes Lunar Radio Cherenkov Experiment, followed by the GLUE experiment at the Goldstone Observatory in California, both searched for neutrino interactions in the 
moon's regolith~\cite{glue,James:2007rq,Bray:2014dva}.   
RICE was an array of radio antennas that was deployed on strings of AMANDA (the predecessor of IceCube), 
ran from 1999-2011, and
published competitive limits as can be seen in Figure~\ref{fig:limit}~\cite{rice}. 

\subsubsection{Experiments Using the Air Shower Technique}
There are also experimental efforts to demonstrate techniques required to detect radio
emission from extended air showers induced by neutrino interactions.  Tau neutrinos that undergo a charged current interaction 
in the Earth and are moving at a slight up-going angle or interact in a mountain 
produce a tau lepton that emerges from the Earth's surface and decays in the atmosphere.
The charged particles in the shower(s) resulting from the tau decay generate radio emission in part due to
a geomagnetic effect: 
positively and negatively charged particles are split due to the Earth's magnetic field, yielding geosynchrotron emission.
Askaryan radiation is also emitted from the shower.  Detection of radio emission from cosmic-ray air showers has been 
made by a variety of experiments~\cite{anitaCR,lofar,trend,aera}, but neutrino-induced air showers are 
much more rare and have yet to be detected.
Auger looks for air showers induced by decays from tau leptons that 
might emerge from neutrino interactions in the Andes Mountains~\cite{augerLimit}.
The TREND project was designed as a prototype to demonstrate the technique~\cite{trend,trend2}, and recently the 
Giant Radio Array for Neutrino Detection (GRAND) 
experiment has been proposed to detect radio emission from neutrino-induced extended air showers~\cite{grand}.

\section{Balloon Experiments}
\subsection{ANITA: The Antarctic Impulsive Transient Antenna}
\begin{figure}[ht]
  \centering{\includegraphics[width=0.35\textwidth]{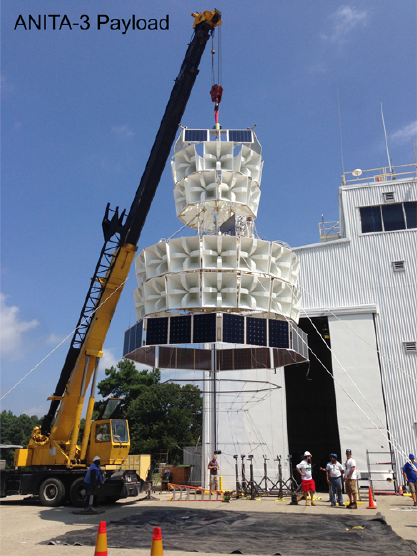}
    \includegraphics[width=0.6\textwidth]{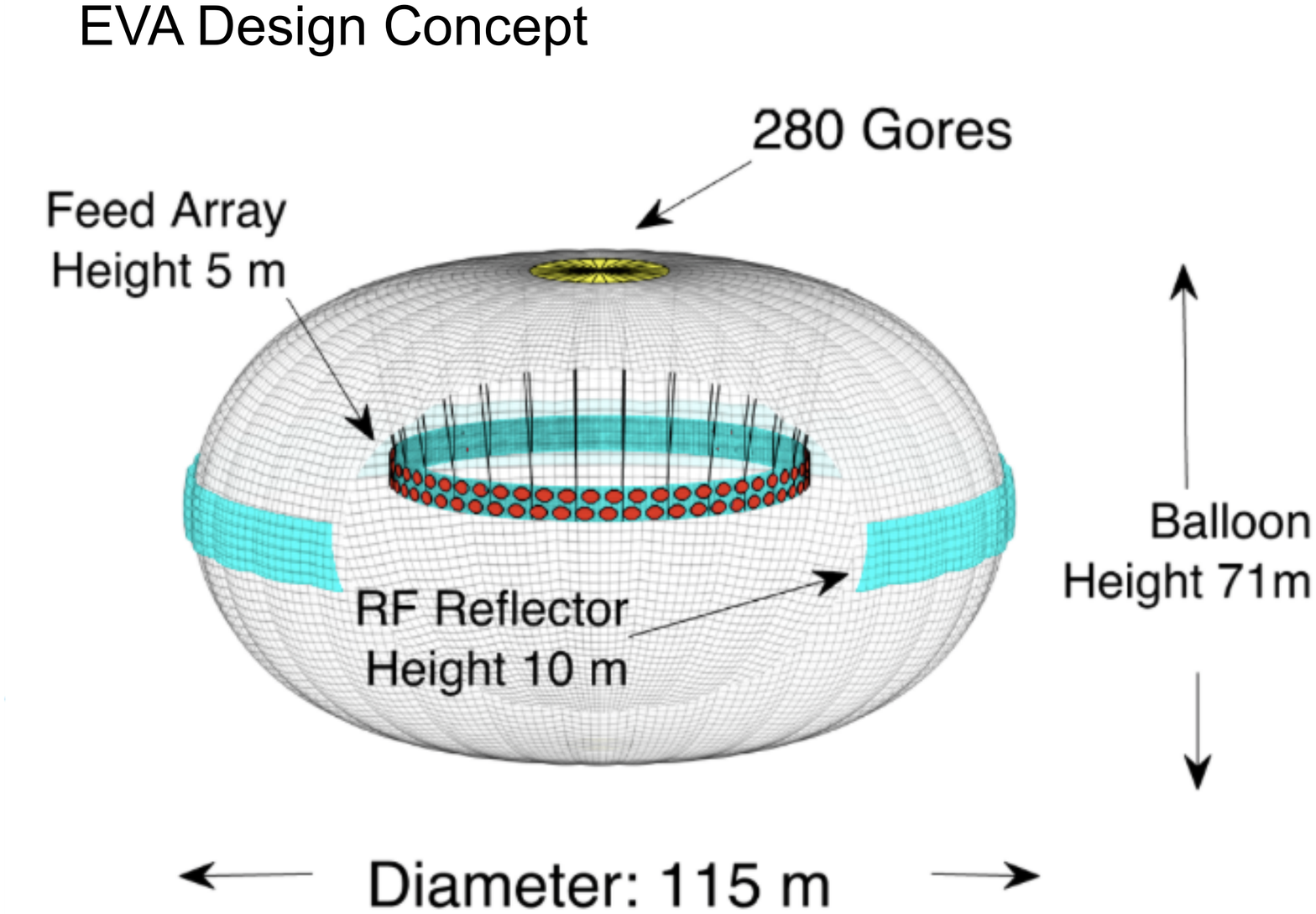}}
  \caption{Left: A picture of the ANITA-3 payload, fully integrated in August 2014
    in preparation for its Antarctic flight in December 2014.  ANITA-3 has 96 channels, as 
    well as a low-frequency antenna for cosmic-ray science (visible below the payload in the picture)
    that was dropped down below the payload after launch. Right: a sketch of the EVA concept, credit to Peter Gorham.}
  \label{fig:anita3eva}
\end{figure} 

The ANITA experiment flies under NASA's Long-Duration Balloon program, lifted to 37~km altitude
over the Antarctic continent by a balloon that is $\sim100$~m in diameter at altitude.  ANITA searches for radio signals from neutrinos interacting in the ice sheet below.
With the horizon at $\sim700$~km distance at float altitude, ANITA uses
all of its visible Antarctic ice sheet as its neutrino detection volume.
The instrument can view $\sim1.5\times10^6$~km$^2$ of ice at altitude, and is designed to search for impulsive, broadband signals predominantly in the vertical polarization 
emerging from the ice below the payload.
ANITA uses dual-polarization (horizontal and vertical), quad-ridged horn antennas with 200-1200~MHz bandwidth.
For each channel (one polarization of one antenna), the signal at the antenna output is amplified and filtered before being split so that it can be sent through
both the trigger and readout chains.
Each channel is Nyquist-sampled and read out with fast ($>2$GSa/sec) 
digitization, producing a $\sim$100~ns waveform for each channel 
when a multi-level trigger is satisfied.  
ANITA-1 was launched in December~2006 
and was aloft for 35~days, ANITA-2 flew in 2008-09 for
31~days, and ANITA-3 flew in 2014 for 22 days.  

There are two main types of triggered events that need to be rejected in order to search for neutrino candidate events:
thermal noise events and human-made noise events.   Thermal noise backgrounds are reduced by requiring a high signal-to-noise
ratio, a high cross-correlation between waveforms with time delays expected from an incoming plane wave, and a linearly polarized
signal.  Events that contain modulated,
continuous-wave (CW) signals are subject to a notch-filter, which removes power in a narrow band surrounding any 
peaks in the measured spectrum above a threshold.

ANITA has published results from searches for a diffuse neutrino flux in each of its first two flights, and also reported
a targeted search for neutrinos from Gamma Ray Bursts (GRBs) using data from its second flight.
ANITA searches for isolated signals from the ice that are not associated with any known bases or with 
any locations with repeating signals.
A blind analysis of data from the second flight of ANITA rejected thermal noise 
by a factor of $\leq 2.5\times 10^{-8}$ and yielded one candidate neutrino 
event on an expected background of $0.97\pm0.42$ events.  This search produced what is still 
the world's best limit on the neutrino flux at energies greater than $10^{19.5}$~eV~\cite{anita2,anita2err}.  
A set of baseline models for cosmogenic neutrino fluxes would predict 0.3-1.0 neutrinos to pass the cuts.
A targeted search for 
neutrinos associated with GRBs was done by narrowing the time window of interest to the 10 minutes surrounding 
12 known GRBs that occurred during the second flight of ANITA when the payload was located in quiet regions of 
Antarctica.  
The strategy of conducting the search
 over a reduced time period leads to reduced backgrounds compared to the diffuse search, allowing 
 for the analysis cuts to be loosened. 
 The two hours of data surrounding each GRB, excluding the 10 minute window of
interest, was used to estimate the background and set the cuts. 
 The GRB search 
yielded no candidate events and the best limit on the 
fluence of neutrinos associated with GRBs at extremely high energies~\cite{anitaGRB}.

ANITA-1 also reported the observation of 16 cosmic rays from geosynchrotron emission.  For 14 of the events,
the impulses were observed after a reflection from the ice surface, and were flipped in polarity 
compared to the remaining two events that were observed directly~\cite{anitaCR}.

ANITA-3 flew 48 antennas compared to 40 in ANITA-2, and triggered on both polarizations (cosmic ray 
air showers are detected in the horizontal polarization~\cite{anitaCR}), with a noise-riding trigger threshold. 
These improvements led to a factor 
of $\sim5$ improvement in expected neutrino event rates compared to ANITA-2.  \mbox{ANITA-3} 
also included a low-frequency
omnidirectional antenna (ALFA), providing additional information for the hundreds of
UHE cosmic-ray events that are expected to have been recorded by ANITA-3 via radio emission from cosmic-ray induced 
atmospheric extended air showers. 
The left-hand panel of Figure~\ref{fig:anita3eva} is a picture of the ANITA-3
payload, fully integrated in August 2014 in advance its December 2014 flight.

\subsection{EVA: The ExaVolt Antenna}  
Building on ANITA's strategy of utilizing NASA's long-duration balloons to
carry out searches for neutrinos in the UHE regime, 
EVA~\cite{Gorham:2011mt} is an ambitious project aiming to transform
a 100~m-scale balloon into a high gain, toroidal reflector antenna (see the right-hand panel of 
Figure~\ref{fig:anita3eva}).  The increase in gain compared to ANITA would mean 
a $\gtrsim100$ times reduction in
threshold in power, and thus a $\gtrsim10$ times reduction in threshold in field strength and neutrino energy.
What makes the EVA concept feasible is the super-pressure balloon (SPB) technology 
currently under development by NASA~\cite{superpressure},
where the inside of the balloon is kept at a higher pressure than the outside ambient pressure, with differential
pressures up to 180~Pa,
keeping the shape of the balloon nearly unchanged over a flight.  
A test flight in 2008 over
Antarctica, SPB flight 591NT, reported that the height and diameter of a 7 Mft$^3$ balloon
changed by 1\% in a 54 day
flight~\cite{pumpkin_balloons}.

\begin{figure}[h]
        \vspace{-1.0in}
          \centerline{\includegraphics[width=13cm]{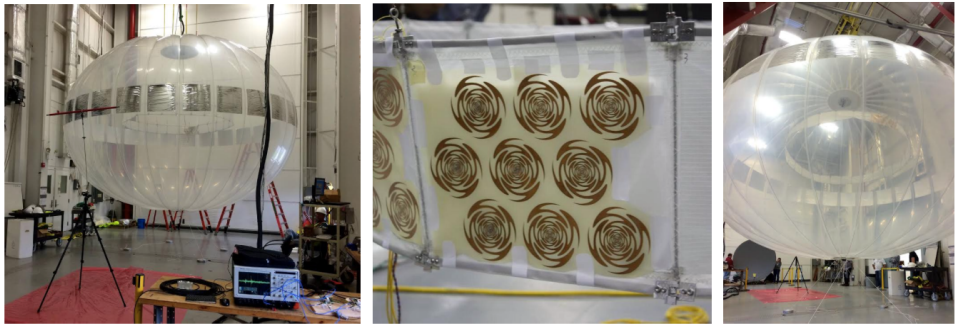}}
               \vspace{-1.0in}
          \caption{Pictures of the EVA prototype test (balloon, toroidal reflector, and dual-polarized 
            sinuous feed antennas) at NASA's Wallops Test Facility~\cite{eva_andresicrc}. Photo credit to Peter Gorham.}
          \label{fig:wallops_test}
        \end{figure}

In September of 2014, the EVA team carried out a test  of a prototype instrumented SPB 
 in a hangar at NASA's Wallops Test Facility 
to demonstrate that a feed array could be deployed inside of
 a SPB, and that the balloon could be instrumented with an RF reflector-receiver system that could be 
 well understood by modeling~\cite{eva_andresicrc}.
 The 1:20 scale prototype was a 5.7~m SPB balloon. 
Reflector tape was attached to the balloon near its equator
 and a feed array membrane held
 the receivers (dual-polarized sinuous patch antennas) 
 and associated electronics over a section of its circumference for the test (see Figure~\ref{fig:wallops_test}).  

 An impulsive plane-wave calibration signal was sent to the balloon and 
was observed at incidence and after reflection.  A comparison
 of the two pulses is consistent with detailed time-domain modeling of the prototype, 
 and the team is currently preparing to propose a flight of a full EVA.

\section{Ground-Based Experiments}
\subsection{Askaryan Effect in Ice}

\subsubsection{ARA: The Askaryan Radio Array}

\begin{figure}[ht]
  \centering{\includegraphics[width=0.5\textwidth]{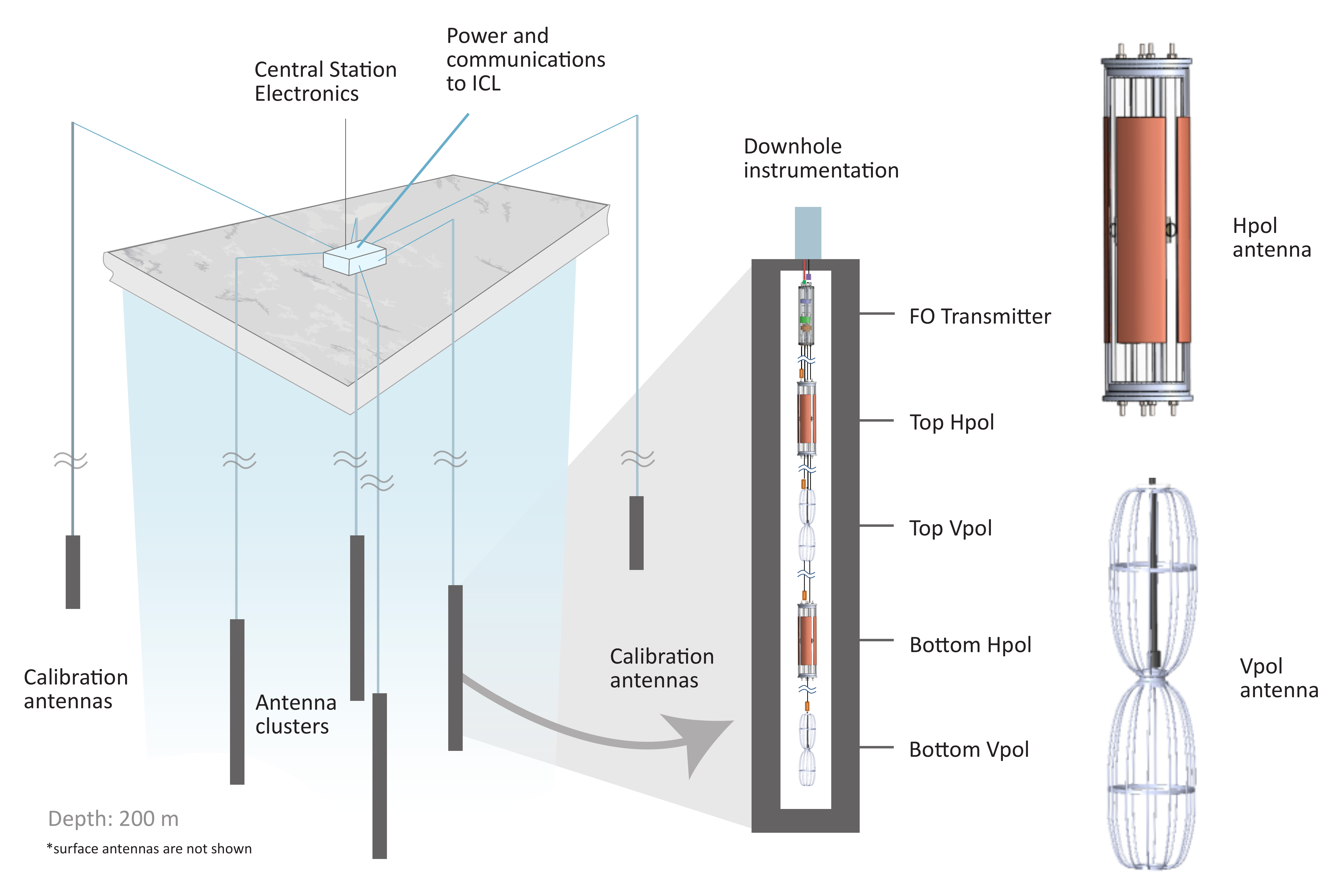}
    \includegraphics[width=0.49\textwidth]{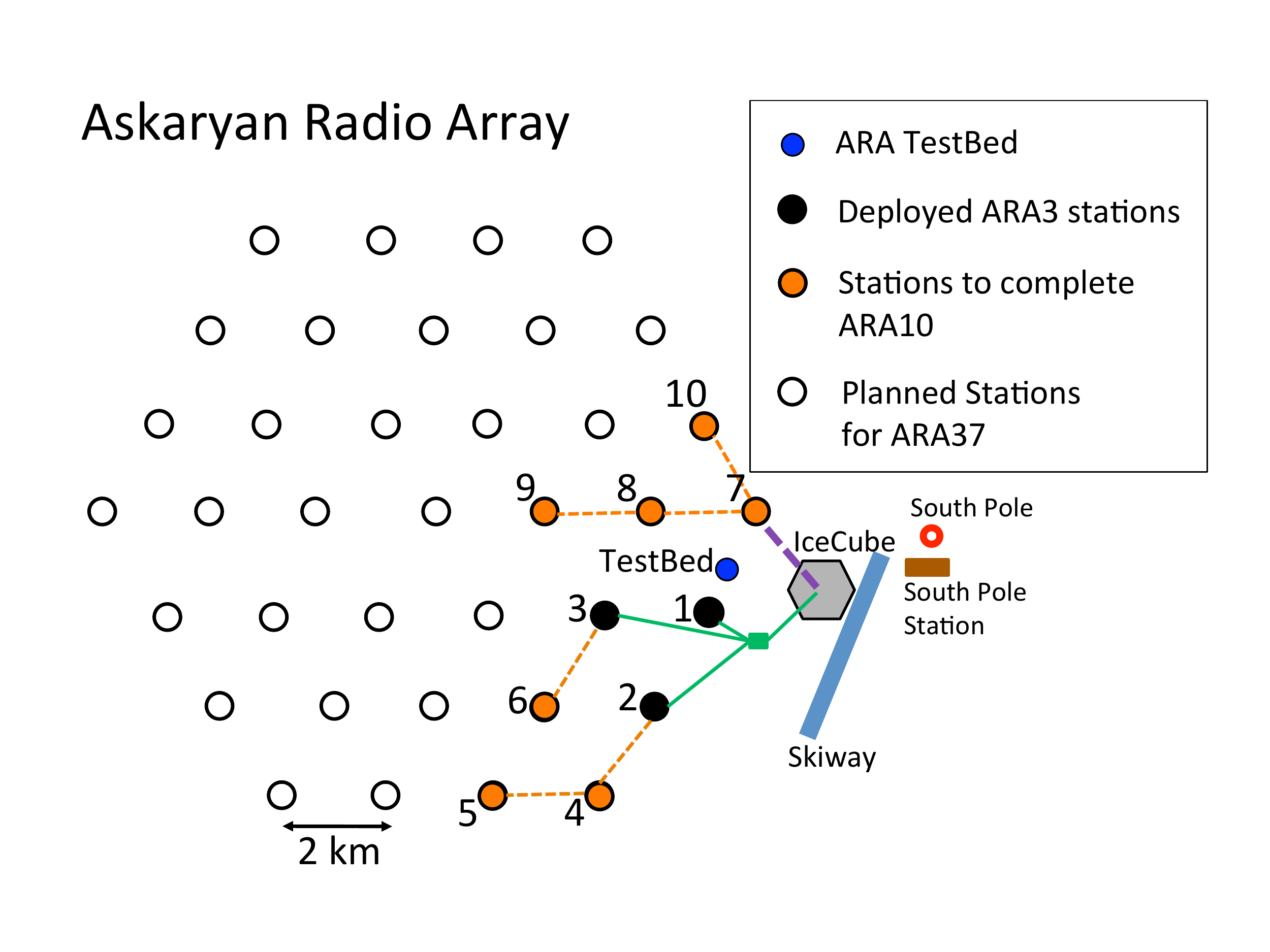}}
  \caption{Left: the baseline design of an ARA station showing the details of one string and a drawing of
  the antennas sensitive to each linear polarization, from Reference~\cite{Allison:2015eky}.  
  Right: a map of the proposed locations
  of ARA stations at the South Pole, from Reference~\cite{Allison:2014kha}. 
  Currently there are three stations deployed (shown in black), plus a testbed 
  (in blue).}
  \label{fig:ara}
\end{figure}

The Askaryan Radio Array (ARA) is a radio array embedded in the ice near the South Pole with the
aim of measuring cosmic neutrinos in the energy regime above $\sim10^{17}$~eV.  
Searching for Askaryan emission from neutrino interactions
in the deep ice near the South Pole, stations of receiver antennas are deployed in dry holes 
at 200~m depth.  A schematic of the ARA design is shown
in Figure~\ref{fig:ara}.  Each station consists of 16~antennas, each of which is read out as a separate channel.
Each ARA station uses a mixture of antennas designed to measure 
vertical and horizontal polarizations, deployed
along four strings separated by $\sim$10~m.  
Calibration pulsers and associated transmitting antennas
are also deployed with each station along additional strings.  

 The trigger and readout chains are similar to ANITA, but the digitized waveforms are $\sim250$~ns in length.
 The trigger currently requires that some number of channels in a station (typically 3 out of 16) exceed a 
 power threshold ($\sim5-6$ times the mean power) within a 110~ns time window, approximately 
the time it takes a signal to traverse a station.
    
  An initial Testbed prototype station was deployed in the 2010-2011 season at $\sim30$~m depth, followed by
  the first full station (A1) in 2011-2012 at 100~m depth.  Two other stations, A2 and A3, have been in the ice
  at 200~m depth since the 2012-2013 season.  
   All three deep stations are currently operational.  Two more stations, A4 and A5, will be deployed in the 2017-2018 season.

ARA neutrino searches look for signals that could have originated from neutrino interactions in the ice and reject events
that reconstruct to the South
  Pole Station.  
 Interferometric techniques, with timing based on cross-correlations between waveforms from different antennas,
 are used to reconstruct directions of signals.
 ARA has the capability to take into account the bending of 
 the trajectories of signals in the firn near the surface, where the index of refraction changes with depth. 
 Thermal noise is reduced through requirements that the timing of the signals in the antennas 
 is consistent with a real signal crossing the detector.

Using the ARA Testbed station, ARA carried out a search for cosmic neutrinos from diffuse sources, and then
a targeted search for neutrinos from GRBs, with 415 days of livetime using the 2011-2012 data 
   from the ARA Testbed~\cite{Allison:2014kha,Allison:2015lnj}.
  As was done for the ANITA GRB search, neutrinos were sought 
in a 10 minute time window surrounding the occurrence of each of 57 GRBs.    
This resulted in the first quasi-diffuse
limit on the GRB flux above $10^{16}$~eV.  The quasi-diffuse limit is based on the assumption that the 57 GRBs considered
for the analysis were representative of all GRBs.

ARA has completed a first search for neutrinos from two of the three deployed deep stations (A2 and A3)
using 10 months of data from 2013~\cite{Allison:2015eky}. 
 No candidates were observed on a background of $0.009 \pm 0.010$ for A2 and $0.011 \pm 0.015$ for A3, resulting
in constraints on the cosmic neutrino flux.

\subsubsection{ARIANNA: The Antarctic Ross Ice Shelf Antenna Neutrino Array}
  
ARIANNA is a proposed ground-based array of radio antennas on the Ross Ice Shelf in 
Antarctica~\cite{Barwick:2014rca}.  An
ARIANNA array of $\sim$1000 autonomously-powered detectors would achieve a sensitivity similar to a 37-station
ARA array.  Although a 1000-station wind-powered ARIANNA design has been considered,  a 1296 ($36 \times 36$)
station array can be powered via solar power and battery backup for the same cost.
ARIANNA's
shallow design with no drilling needed
makes an ARIANNA station simpler to deploy than an ARA station.
ARIANNA's proposed 36$\times$36 station array would span more than 1000~km$^2$ on the $\sim600$~m-deep
Ross Ice Shelf compared to an ARA37 that would span $\sim$100~km$^2$ area at the South Pole where the ice
is $\sim2800$~m deep.
There is a reflecting layer of ocean water at the bottom of the Ross Ice Shelf that acts as a mirror, 
reflecting radio signals from down-going neutrinos back up to the antennas on the surface of the
snow~\cite{fenfang_thesis}.  This means that ARIANNA would have a broader range in visible solid angle 
compared to ARA due to the ability to see down-going neutrinos, which counteracts the shorter measured radio 
attenuation length of the ice on the Ross Ice Shelf.

  \begin{wrapfigure}{r}{0.50\textwidth}
  \centering{\includegraphics[width=0.45\textwidth]{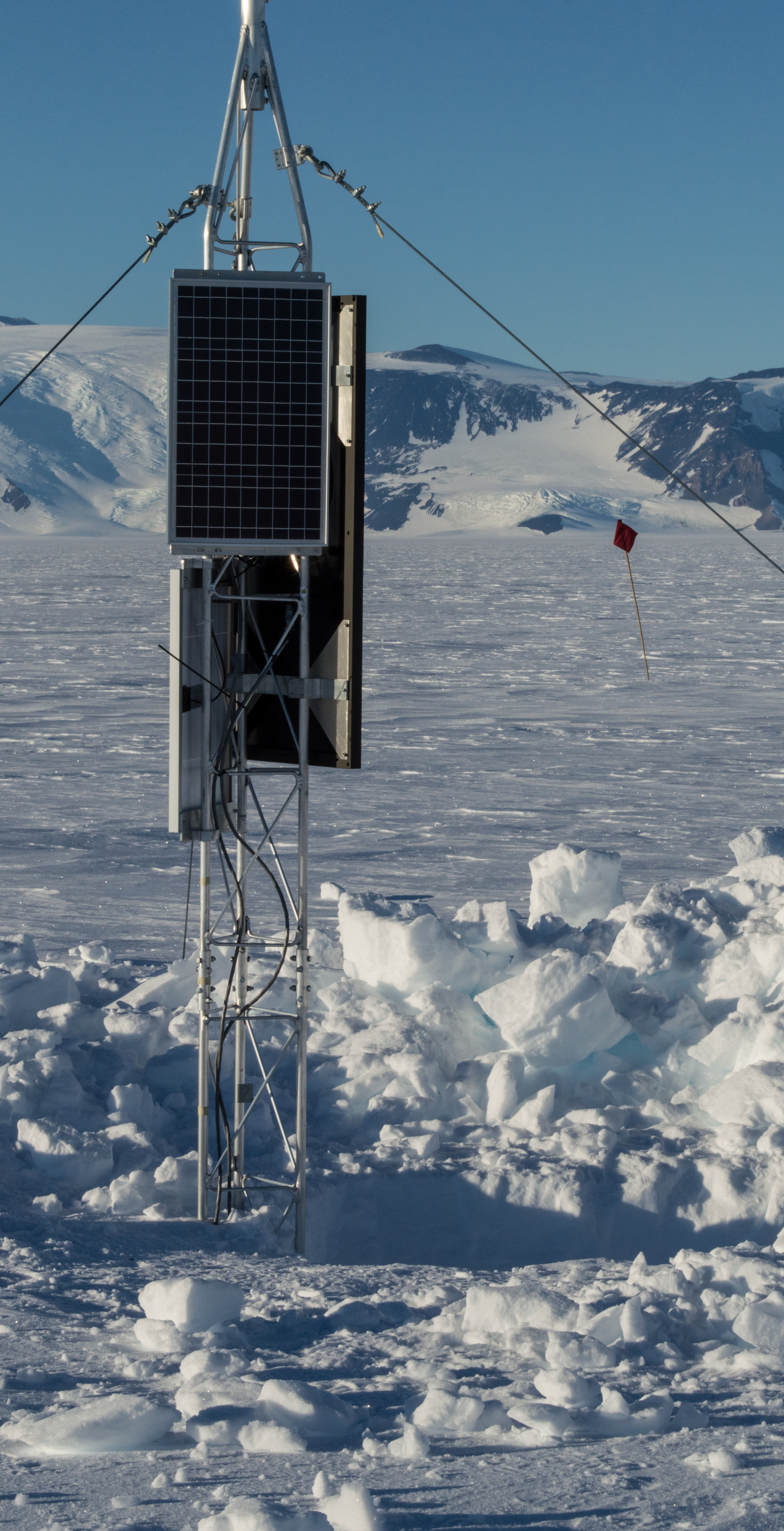}}
  \captionsetup{width=0.45\textwidth}
  \caption{View of one ARIANNA station with solar panels on a tower.   The box containing the station's data acquisition box is buried in the ice beneath the tower.  Photo credit to Chris Persichilli.}
  \label{fig:arianna}
\end{wrapfigure}

An ARIANNA station, shown in Figure~\ref{fig:arianna},
consists of high-gain log-periodic dipole antennas (LPDAs) with a bandwidth of $\sim0.08-1.3$~GHz in snow 
that are downward-pointing and deployed near the surface of the snow.   The HRA (Hexagonal Radio Array) consists 
of eight prototype stations, most consisting of four LPDAs, 
and was deployed in the Austral summers of 2012-2013 and 2014-2015.   
A design station for the full ARIANNA array would contain 6-8 downward-facing LPDAs. 
When a multi-level trigger is satisfied at a station, waveforms across that station are
digitized at 2~GSa/sec.  
Stations are solar powered, with battery backups that both provide power on cloudy days and extend the life of the 
stations a few weeks past sundown in low power mode, as discussed in 
Reference~\cite{Kleinfelder:2013zya}.  The achieved livetime is 58\% of the year~\cite{Barwick_Berg_Besson_Binder_2015}.

ARIANNA performed a first neutrino search using data from the HRA.  After searching for events that contain 
impulsive signals
whose arrival times are consistent with a plane wave crossing the station in the absence of any
narrow peak in the Fourier spectrum
pointing to CW noise, no neutrino candidates were found in four months of operation. 
From those results,
ARIANNA set its first limit on the diffuse neutrino flux above $10^{17}$~eV with three HRA stations~\cite{Barwick:2014pca}.
Data from all eight stations of the HRA is currently being analyzed.

\subsubsection{GNO: The Greenland Neutrino Observatory}

An effort is ongoing to develop Summit Station, a year-round station run by the National 
Science Foundation (NSF), as a site for radio detection of UHE neutrinos.
The ice at Summit Station is comparable to that at the South Pole in radio clarity (see Section~\ref{sec:attenuation_ice}) 
and in thickness (3000~m at Summit Station compared to 2800 at South Pole), but with a shallower firn allowing for shallower deployment
of sub-surface antennas.  
At Summit Station, at 100 m depth the  firn density has reached 95\% that of deep ice, 
compared to 140 m at the South Pole~\cite{Alley_Koci_1988,Kuivinen_1983}.
A first prototype station for GNO was deployed
in the Summer of 2015 to monitor the radio-frequency noise environment.
      
\subsection{Lunar Experiments}      
        
        In 1995, the Parkes Lunar Radio Cherenkov Experiment, using the 64~m Parkes radio telescope, was the first to search for radio emission from neutrino
        interactions in the moon's regolith, the loose, $\sim$10~m deep layer at the surface of the moon.
         Due to the expected geometry of neutrino-induced events, lunar radio 
         experiments are expected to be sensitive to
         neutrinos when viewing the lunar ``limb'' (the edge of the moon's visible surface)~\cite{braythe2016}.
         Multiple antennas are beamed for heightened sensitivity to regions along the limb. 
          Radio frequency interference is often reduced in lunar radio experiments 
          by requiring signals in different frequency bands to 
          arrive at the delays expected from dispersion through the ionosphere.
         GLUE was the next to search for highly energetic neutrino interactions in the 
         moon's regolith in 2000-2003 using the 34~m DSS13 and 70~m DSS14 antennas at the Goldstone 
         Observatory in California for 124~h of observing time, constraining the neutrino flux at energies 
         above $10^{21}$~eV~\cite{James:2007rq,glue}. 
         Since then, a series of experiments have followed up with lunar observations with radio 
         telescopes~\cite{beresnyak:2005,jameslunaska2010,jamesultra-high2009,Jaeger:2009zb,spencerla2010,Bray:2014dva,Bray_Ekers_Roberts_Reynolds_2015}.
         The NuMoon experiment observed the moon with the Westerbork Synthesis Radio Telescope array in four frequency
         bands between 113 and 175~MHz for 47.6~hours, and
         claims the most competitive constraints on the neutrino flux above $10^{23}$~eV, as seen in 
         Figure~\ref{fig:limit}~\cite{buitinkconstraints2010}.  Lunar observations have been proposed with the 
         Low Frequency Array (LOFAR), a recently completed radio antenna array with beaming accomplished electronically
         rather than mechanically, with a sensitivity projected to improve upon the NuMoon constraints by a factor of 25
         as stated in Reference~\cite{buitinkconstraints2010}.   The Square Kilometer Array (SKA) will be even more sensitive
         to neutrinos from the moon with this technique due to its wide bandwidth and large collecting area.  Construction of
         Phase~1 of SKA is scheduled to begin in 2018~\cite{Bray_Alvarez-Muniz_Buitink_2014}.
         
\subsection{Neutrino-Induced Air Showers}
\label{sec:air_showers}
Auger searches for neutrinos in the UHE regime 
by distinguishing ``young'' and ``old'' showers using their 1600 Surface Detectors,
each consisting of a Cherenkov tank with photomultipliers read out with 
flash analog to digital converters (FADC) with 25~ns resolution~\cite{augerLimit}.  
Protons, heavy nuclei and any
gamma rays are expected to shower soon after they hit the atmosphere, so by the time their showers hit
Auger's Surface Detectors they are old showers, consisting mainly of muons with highly coincident arrival times.
Neutrinos however, are equally likely to interact anywhere along their path through the atmosphere, and therefore
can be young when they arrive at the Surface Detectors.  Also, a tau neutrino can interact and produce a tau
lepton in mountains surrounding the observatory and the subsequent tau decay would lead to a young shower seen 
by Auger.
Younger showers would still contain an electromagnetic component when they are observed
and would show a broader distribution of arrival times. 
The greatest separation between young and old showers
is expected for events that are highly inclined.  Auger sets competitive limits on the tau neutrino diffuse flux only,
to which they are the most sensitive.  Auger's competitive UHE neutrino flux limit
shown in Figure~\ref{fig:limit} is a tau neutrino limit multiplied (i.e. weakened) 
by a factor of 3 for comparison with other experiments
whose limits are averaged over all flavors.

The proposed GRAND experiment would consist of an array of antennas in 
a remote mountainous site to search for air showers induced by the decay of tau leptons in the atmosphere
that originate from a charged current interaction of
tau neutrinos with the Earth~\cite{grand}.  The GRAND collaboration proposes the deployment of 
$\mathcal{O}(10^{5})$ radio antennas, operating between $30-100$~MHz and
covering $\mathcal{O}(10^5)$~km$^2$ at a site in the Tianshan mountains in China.  
The preliminary design includes several sub-arrays, each with area $\mathcal{O}(10^4)$~km$^2$.
Initial estimates predict that a $6\times10^4$~km$^2$ array would have an effective area several times
larger than ARA's effective area between $10^{18}-10^{19}$~eV, and a $2\times10^{5}$~km$^2$ array
would further improve the sensitivity by a factor of $\sim$10 in the same energy region.
       
\section{Goals of Future Experimental Efforts}
The most important aim in this field at present is to 
 make a first discovery of neutrinos in the UHE regime.  Alongside the firm expectation of the
 BZ neutrino flux, there is
 an expected neutrino flux from the astrophysical sources themselves that may be comparable.
 Once the initial discovery is made, it will open up a unique window to the universe at the highest energies.
 We will likely be able to say that the highest energy cosmic rays are not purely
 heavy composition, and that no new physics prohibits a UHE neutrino from reaching us from cosmic distances.
We will know the detector effective area  that is needed to measure a large enough sample of neutrinos to carry out
a particle physics and astrophysics program in this new frontier.
In addition, a first discovery will heighten the urgency to develop the tools to reconstruct neutrino directions
and measure neutrino energies, which will be necessary as physics priorities are pursued beyond the implications of
the first detection.  
 
\begin{wrapfigure}{r}{0.65\textwidth}
          \begin{center}\includegraphics[width=0.60\textwidth]{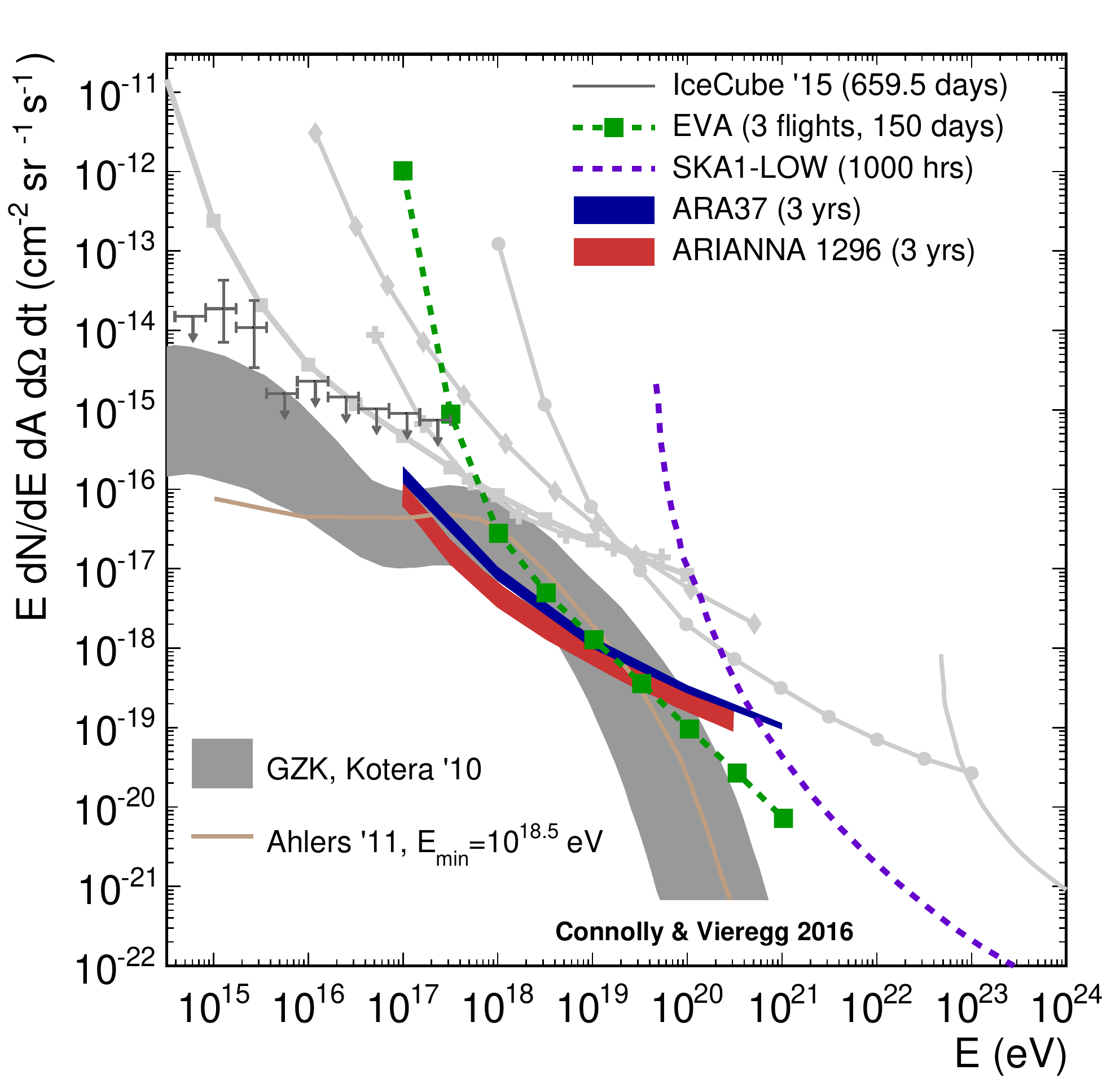}
           \end{center}
           \captionsetup{width=0.60\textwidth}
          \caption{The sensitivity reach of future radio neutrino experiments at ultra-high energies and above.  
             We show projected limits for ARA37, EVA and SKA Phase 1 with the low frequency 
            antennas~\cite{Allison:2014kha,eva_andresicrc,Bray_Alvarez-Muniz_Buitink_2014}.  
            In light gray we also show the current most competitive limits
            that were shown in Figure~\ref{fig:limit} for comparison.
          }
          \label{fig:limit_future}
        \end{wrapfigure}

Both view-from-a-distance and embedded experiments are working to reduce their energy thresholds, which would move their 
sensitivities to the {\it left} in Figure~\ref{fig:limit_future},  
and increase their chance of a discovery with a falling neutrino spectrum.  
In this figure, one can see the effect of EVA's factor 
 of $\gtrsim10$ reduction in energy threshold compared to ANITA.
Many experiments
are working to implement the interferometric techniques that have been successful at the analysis stage to their
trigger stage, to enable detection of lower energy events~\cite{Romero-Wolf:2015,Vieregg:2015baa}.  
With an interferometric phased array trigger
coherently summing hundreds of antennas,
in-ice experiments may be able to lower
their thresholds enough to reach
the higher end of the PeV neutrino spectrum observed by IceCube~\cite{bertErnie,bigBird} and ascertain
whether the spectrum cuts off or continues to higher energies~\cite{Vieregg:2015baa}.  

Experiments are also working to increase their effective area in order to move their sensitivities {\it down} in 
Figure~\ref{fig:limit_future}.  Ground-based projects ARA and ARIANNA are working to expand their
detectors, adding as many 
stations as funding and logistics allow.  The GRAND experiment described in Section~\ref{sec:air_showers} is looking to utilize a vast swath of
the atmosphere to seek neutrino-induced extended air showers and will compete with ARA in the same energy
range.

The radio technique for neutrino detection allows experimenters to instrument the immense volumes 
needed for sensitivity to the rare fluxes of neutrinos expected in the UHE regime through a variety of
projects that view the detection medium from within, with embedded detectors such as ARA or ARIANNA, 
or from a distance, with balloon experiments such as ANITA.
Radio telescopes pointed at the moon such as NuMoon are the most sensitive at even higher energies.
Other projects, such as Auger, search for UHE neutrinos through an air shower signature.
Radio techniques
provide an opportunity 
for a long-term UHE astrophysics 
program that can reach even some of the most pessimistic predictions for the neutrino flux in the UHE regime.  
The results from the first decade and a half of 
neutrino searches in this field and the projections for future experiments portend an exciting future.

\section{Acknowledgements}
The authors would like to thank Tom Gaisser and Albrecht Karle for inviting us to write this contribution.
We are also grateful to Jordan Hanson, Clancy James, Andrew Romero-Wolf and David Saltzberg for helpful feedback.  
A. Connolly would also like to thank the National Science Foundation for their support through CAREER award 1255557, 
BIGDATA Grant 1250720, Grant 1404266 for ARA support, and NASA for their support through 
Grant NNX12AC55G for EVA development.  A.~Connolly would also like to thank
the United States-Israel Binational Science Foundation for their support through Grant 2012077.
A.~Connolly and A.~Vieregg would like to thank NASA for their support for ANITA through Grant NNX15AC20G.
We are grateful to the U.S. National Science Foundation-Office of Polar Programs.  A. Vieregg would like to thank
the Kavli Institute for Cosmological Physics for their support, and A. Connolly would like
to thank the Ohio State University for their support. 

\bibliographystyle{ws-rv-van}
\bibliography{radio}
\end{document}